\documentclass{aa520}

\usepackage{graphicx}
\usepackage{txfonts}

\usepackage{natbib}
\bibpunct{(}{)}{;}{a}{}{,}

\newcommand{\cab}{CAB}

\let\AAA=\AA
\renewcommand{\AA}{\mbox{\AAA}}

\newcommand{\he}{\object{HE~0512--3329}}

\newcommand{\funit}{\mathrm{erg \, s^{-1} \, cm^{-2} \, \AA{}^{-1}}}

\newcommand{\mumi}{\mathrm{\muup m^{-1}}}

\newcommand{\sub}[1]{_\mathrm{#1}}

\newcommand{\supp}[1]{^\mathrm{#1}}

\newcommand{\rtext}[1]{\quad \mbox{#1}}

\newcommand{\kmsmpc}{\mathrm{km\,s^{-1}\,Mpc^{-1}}}

\begin{document}

\title{Disentangling microlensing and differential extinction in the
  \\   double QSO \he}

   \author{O. Wucknitz
          \inst{1}
          \and
          L. Wisotzki\inst{1,2}
          \and
          S. Lopez\inst{3}
          \and
          M.~D. Gregg\inst{4}
          }

   \offprints{O. Wucknitz}

   \institute{Institut f\"ur Physik, Universit\"at Potsdam, Am Neuen Palais,
     D-14469 Potsdam, Germany\\ \email{olaf@astro.physik.uni-potsdam.de}
     \and
     Astrophysikalisches Institut Potsdam, An der Sternwarte 16, D-14482
     Potsdam, Germany 
     \and
     Departamento de Astronom\'ia, Universidad de Chile, Casilla 36-D,
     Santiago, Chile 
     \and
     Institute of Geophysics and Planetary Physics, Lawrence Livermore
     National Laboratory, 7000 East Ave, L-413, Livermore, CA 94551-9900
     }

   \date{Received 29 January 2003 ; accepted 14 April 2003}

   \abstract{
     We present the first separate spectra of both components of the
     small-separation double QSO
     \he\ obtained with HST/STIS in the optical and near UV. The similarities
     especially of the emission line profiles and redshifts
     strongly suggest
     that this system really consists of two lensed images of one and
     the same source.
     
     The emission line flux ratios are assumed to be unaffected by
     microlensing and are used to study the differential extinction effects
     caused by the lensing galaxy. Fits of empirical laws show that the
     extinction properties seem to be different on both lines of sight. With
     our new results, \he\ becomes one of the few extragalactic systems which
     show the 2175\,\AA\ absorption feature, although the detection is only
     marginal.
     
     We then correct the continuum flux ratio for extinction to obtain the
     differential microlensing signal. Since this may still be significantly
     affected by variability and time-delay effects, no detailled analysis of
     the microlensing is possible at the moment.
     
     This is the first time that differential extinction and microlensing
     could be separated unambiguously. We show that, at least in \he, both
     effects contribute significantly to the spectral differences and one
     cannot be analysed without taking into account the other.  For lens
     modelling purposes, the flux ratios can only be used after correcting for
     both effects.
     
      }

   \maketitle

\section{Introduction}

\he\ was discovered as a probable lensed quasar in the course of a snapshot
survey with the Space 
Telescope Imaging Spectrograph (STIS). It is a doubly imaged QSO
with a source redshift of $z=1.58$ and an image separation of
$0\farcs644$. The lensing galaxy has not been detected 
yet, but 
strong metal absorption lines with a redshift of $z=0.93$ identified in the
integrated spectrum provide good evidence
that a damped Ly$\alpha$ (DLA) system intervenes at this redshift. This system
is 
very probably associated with the lensing galaxy \citep{gregg00}.

The absorption lines provide the possibility to study the
interstellar matter (ISM) of the lensing galaxy
along the two lines of sight,
separated by 5.1\,kpc at the proposed lens redshift,\footnote{We assume a low-density flat universe with $\Omega=0.3$,
  $\lambda=0.7$ and $H_0=70\,\kmsmpc$ throughout this paper.} 
 which allows us
to learn more about the nature of the DLA absorber by comparing column
densities of hydrogen and metals as well as kinematic profiles 
along both lines of
sight. \he\ is especially well suited for such investigations because of the
wealth of absorption features and the close image separation.

The propagation effects of microlensing and
extinction in the lensing galaxy can be studied by using the fact that we see
two images \emph{of 
  the same source}. In singly imaged sources, it is not possible to uniquely
identify microlensing or extinction effects because the true source
fluxes and spectra are not a priori known. In the case of multiply
imaged sources, this problem can be circumvented by using only differences of
the individual images as diagnostic. These differences are independent of the
true source spectrum and flux and therefore allow the examination of
differential propagation effects.

An additional complication arises from possible variability of the
  source. Because of the time-delay, observations of both components taken at
  the same time are really snapshots of the source at different epochs. To
  correct for this effect (which is thought to be negligible for the emission
  lines because of the longer variability time scale), (spectro)photometric
  monitoring is necessary to determine the time-delay and obtain measurements
  at the same source epoch.

It has been known from earlier photometric observations that the flux ratio
A$/$B in \he\ shows a strong dependence on wavelength. In the $R$ and $I$
bands, A is brighter 
than B by about 0.45\,mag while the two are almost equal in $B$
\citep{gregg00}. 
A natural explanation for this effect is differential reddening caused by
different extinction effects in the two lines of sight. This scenario is
supported by the strong metal lines already visible in the integrated spectrum
of \he\ \citep{gregg00} which make the existence of significant
amounts of dust very likely.
Under this assumption, \citet{gregg00} estimated a relative
visual extinction of $A_V=0.34\,\mathrm{mag}$ by fitting galactic
extinction curves with $R_V=E(B-V)/A_V=3.1$ on both lines of sight to the
photometric data.
We will see later that this approach is not appropriate in the case of \he.

The other possible explanation is microlensing by stars and other compact
objects in the lensing galaxy. Since the lines of sight cross different star
fields in the lens, the microlensing effect is independent for A and B and can
be observed in the difference.
The lensing effect itself is an achromatic process, but chromatic effects are
nevertheless expected if different parts of the source, which are influenced
by different microlensing amplifications, have different colours.
Since the microlensing amplification pattern has a high contrast on very
small scales, the effectively observed total amplification of the integrated
source depends significantly on the size of the source.
The effect is maximal for point sources but is smoothed out and thus
reduced for larger sources.

Standard models of QSOs consist of different emitting regions of very
different sizes \citep[see e.g.][]{krolik99}. Smallest is the continuum
region, which itself scales with 
wavelength. Broad line regions are expected to be much larger
than the continuum region but smaller than the narrow line regions.
The whole range of scale lengths
covers several orders of magnitude.
This means that the microlensing signal should be strongest for the continuum,
only very weak at most for the broad emission lines and absent for narrow
lines. 

This size-dependent behaviour of microlensing gives a possible clue to
separate the 
effect from differential extinction and study the two processes separately.
Microlensing is smeared out in the emission lines
whose fluxes are thus only influenced by the differential extinction. The
continuum, on the other hand, is also affected by microlensing. The extinction
itself is a smooth function of wavelength but does not depend on the
source size.
We can therefore
use the emission lines to study the extinction directly without worrying about
microlensing and correct the continuum for extinction to study the pure
differential microlensing in the remaining differences.

The spectra we use to follow this approach
were obtained with
STIS on board the HST in the optical and near UV covering the QSO emission
lines of \ion{Ly}{$\beta$}, \ion{Ly}{$\alpha$}, \ion{Si}{iv}, \ion{C}{iv} and
\ion{C}{iii]} with the goal to study the differential extinction by comparing
the continua of the A and B images.

The results of these observations are presented in this paper. We show
the first \emph{separate} spectra of the A and B images and use the similarity of
these to confirm the nature of the object as a gravitational lens system.
We use the combination of continuum and emission lines to disentangle the
effects of extinction and microlensing as described above.
We see 
that microlensing actually does
influence the flux ratio significantly and that the continuum cannot be
used to study the extinction directly.

To analyse the differential extinction we will fit empirical
extinction laws to our emission line data and discuss the results.
The microlensing data will be used to estimate an upper limit of the
source size using the assumption that variability and time-delay effects can
be neglected.

This is the first time that such a detailed attempt to
separate the effects of differential extinction and microlensing has been
successfully performed for any lens system.

\begin{figure*}
\centering
\includegraphics[angle=90,height=0.9\textheight]{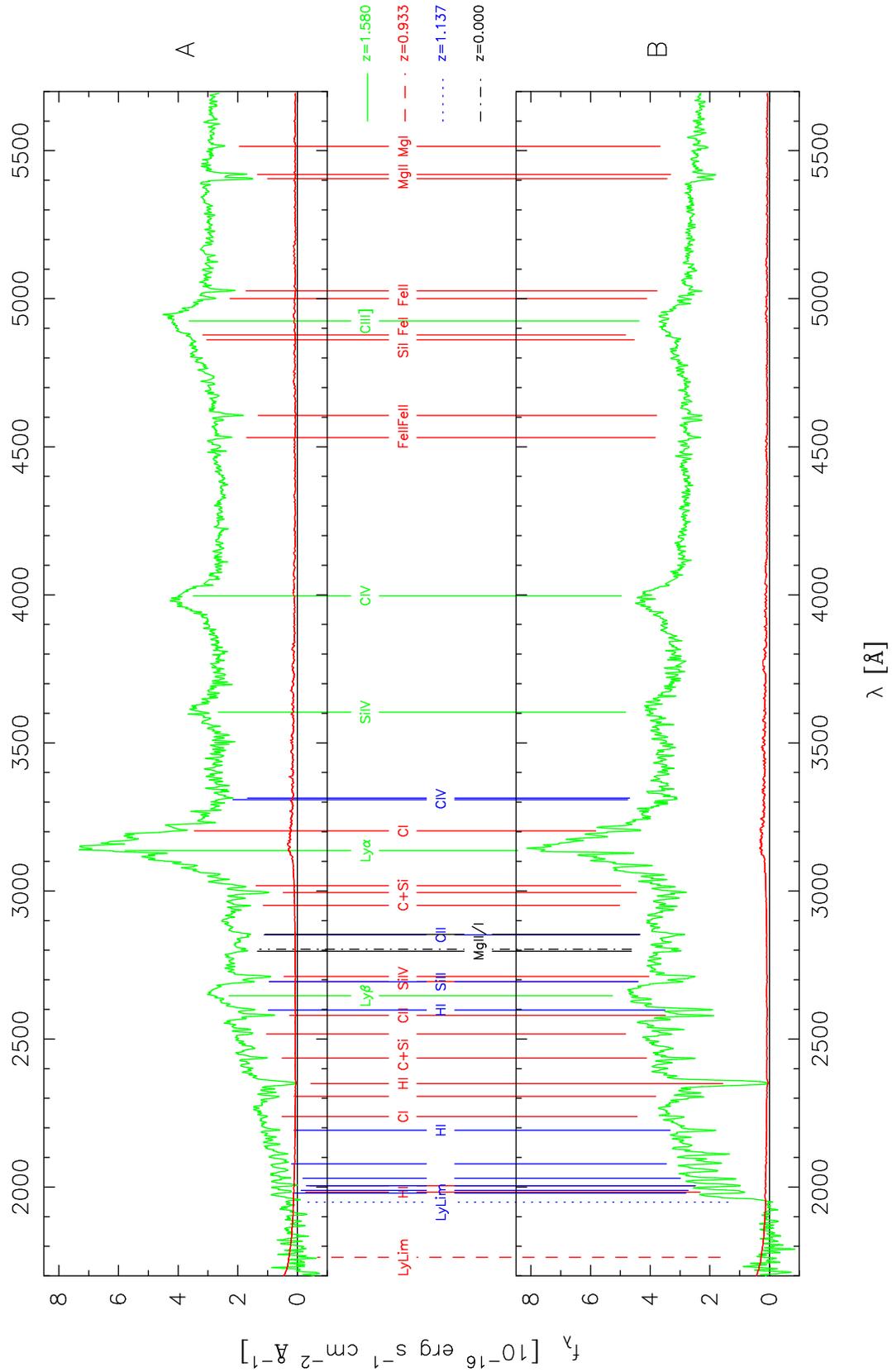}
\caption{The combined spectra of \he, smoothed with a Gaussian filter of
  FWHM=$3.5$\,\AA.  Component A is at top, B at bottom. The
  curve near $f_\lambda=0$ shows the statistical errors. The
  middle panel marks emission and absorption features at $z=1.58$ (QSO
  emission), $z=0.933$ (DLA system), $z=1.137$ (Lyman limit system) and $z=0$
  (galactic absorption).
}
\label{fig:plotspec}
\end{figure*}

\section{Observations and initial data reduction}

To cover a sufficiently large spectral range, we used the CCD for the
optical and the near UV MAMA detector for the UV part. The slit was
positioned along the line connecting the two QSO components.

The CCD spectra were taken on 13 August 2001 with the G430L grating and 52x0.5
aperture, covering a 
spectral range from 2900 to $5700\,\AA$ 
\citep{stis_instr}.  With a spectral
element size of $2.73\,\AA$, the resolution is about $3.8\,\AA$. 
Three exposures with total integration time of 2266\,sec and a standard
dithering pattern were used.

Data reduction started with a standard
2-dimensional reduction including flat-fielding, flux and
wavelength calibration with the CALSTIS software Version 2.10
\citep{stis_data} in IRAF V2.12 with STSDAS Version 2.3.
Cosmic ray rejection could not be performed in a
standard way because the individual exposures were not done in CR-SPLIT mode. 
We therefore used our own software, developed specifically for this task,
to apply the shifts, reject cosmic ray events and combine the images.

We then used the CALSTIS task X1D to extract the spectra of both QSO
components separately. A smaller than standard extraction width was used to
minimize the fraction of measurements being rejected due to flagged bad pixels
or cosmic rays.

The three resulting extracted spectra were finally combined and cleaned
of still remaining cosmic rays and otherwise deviant pixels with our own
software.

The MAMA spectra were taken with the G230L grating and 52x0.2 aperture, 
covering a range from 1570 to $3180\,\AA$ with an element size of $1.58\,\AA$
and a resolution of $3.3\,\AA$.
The six exposures of a total of 16588\,sec were taken on 15 August 2001,
only two days after the CCD spectra. Since the MAMA detector is not sensitive
to cosmics, a standard CALSTIS reduction could be used to calibrate the
2-dimensional images. To extract the two QSO components separately, X1D was
used with a slightly smaller than standard extraction width and with background
regions adapted to the situation of two closely separated but well resolved
spectra. 
The six MAMA spectra were then combined with the same algorithm as the CCD
spectra. 

For the analysis, we did not use the short wavelength end of the combined CCD
spectrum ($<3020\,\AA$) and the long wavelength end of the MAMA spectrum
($>3120\,\AA$). In these regions, the $S/N$ is not
optimal and calibration uncertainties are expected to be worst.
In the remaining overlapping region, the two spectra agree very well and a
simple averaging combination was used for the final spectrum.

The typical error per spectral element of the final combined spectrum is
1--$2\cdot10^{-17}\,\funit$, except near $\lambda\approx
3100\,\AA$ and for $\lambda<1700\,\AA$.
This is equivalent to a $S/N$ of 30 per element in the optical and about 15 in
the UV region.

\section{The spectra}

\subsection{First view}

\begin{figure*}
\hspace{1em}\includegraphics[scale=0.89]{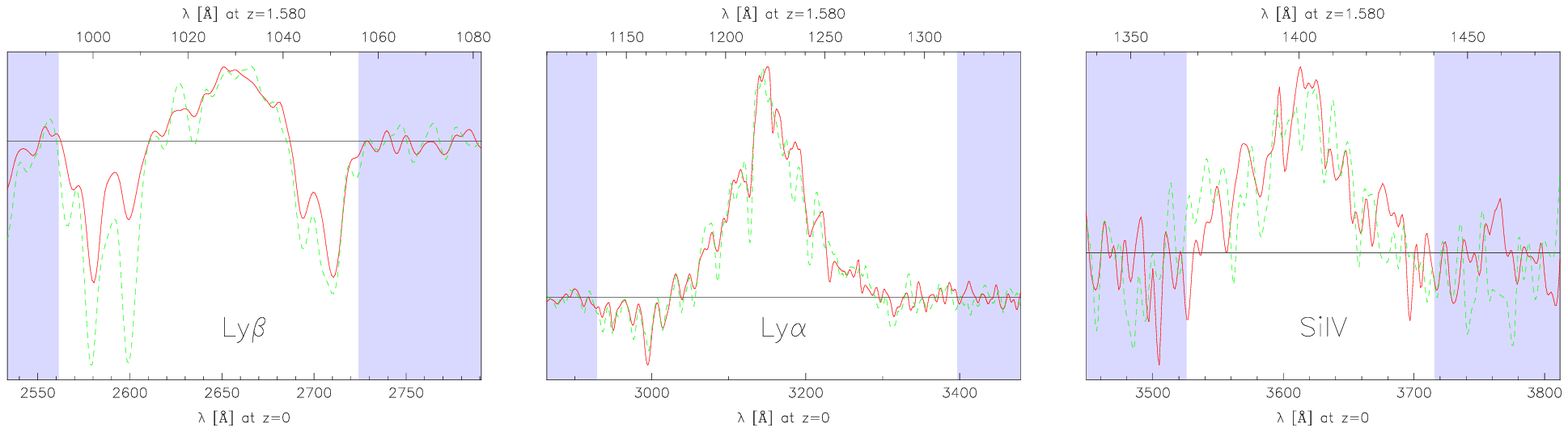}

\vspace{2ex}

\hspace{1em}\includegraphics[scale=0.89]{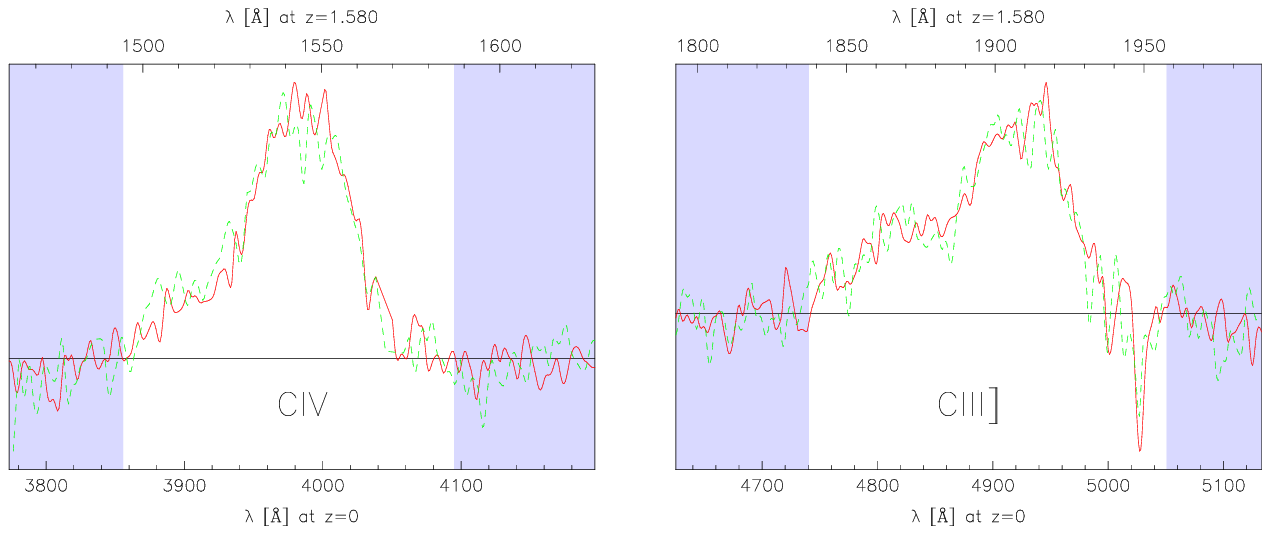}\hspace{1em}
\caption{Profiles of the five emission lines after local continuum
  subtraction. A is plotted solid, B is dashed. The flux of B was scaled by the
  determined emission line flux ratio to obtain the same level for both
  components. The shaded regions were used for the linear continuum fit with
  rejection of absorption features.
  The spectra are smoothed with a Gaussian kernel of FWHM=5\,\AA.
} 
\label{fig:emlines}
\end{figure*}

Figure~\ref{fig:plotspec} shows the combined spectra from all CCD and MAMA
exposures.
These comprise the first separate spectra of the two QSO components in \he.
The
general features of A and B look very similar. The relative strength of the
emission lines, their profiles (see also
Fig.~\ref{fig:emlines} and discussion below) and redshifts agree very
well which strongly
confirms the interpretation as two lensed images of one source. 
The continua, on the other hand, show similarities as well as significant
differences which will be discussed in detail in the later parts of this paper.

The wealth of absorption features seen in both images can provide important
information about column densities of both hydrogen and metals and will be
discussed in detail in another publication. The spectra confirm the existence
of a DLA system at a redshift of $z=0.933$ which was
already 
proposed by \citet{gregg00} to explain the strong low-ionization metal lines
in the 
(then still combined A+B) spectrum. Our estimate of the redshift differs
only slightly from the former measurement by \citet{gregg00} of
$z=0.9313\pm0.0005$. 
Figure~\ref{fig:plotspec} identifies the anticipated DLA
line as well as a high number of metal absorption features at the same
redshift in both components.
The break at about $2000\,\AA$, which is especially prominent in B due to the
higher continuum there, does \emph{not} correspond to this already known
system at $z=0.93$ but is caused by a second intervening Lyman limit system at
a redshift of about $z=1.137$. This system is much weaker than the lower
redshift one but nevertheless shows some metal lines which are also identified
in Fig.~\ref{fig:plotspec}. Weak \ion{Mg}{ii} absorption at a very similar
redshift ($z=1.1346$) was already noticed by \citet{gregg00}.
Finally we detect galactic \ion{Mg}{i+ii} absorption at a redshift consistent
with zero.

A quantitative analysis of the absorption features, combined with
high resolution UVES spectra, will be published in an upcoming paper (Lopez et
al., in prep.). Especially important is the analysis of the DLA lines in our
spectra which provide the \ion{H}{i} column 
density needed to determine metalicities in both lines of sight.

\subsection{Differences in the global shape of A and B}

Although the two 
spectra have many features in common, there are
some highly significant differences.
Most prominent is the very different behaviour of the continuum flux on large
wavelength scales. On the red side of $\lambda\approx 4600\,\AA$, the A
component is brighter than B. For smaller wavelengths, 
especially close to the limit near 2000\,\AA, B becomes much brighter than
A. This effect has already been described by \citet{gregg00} using photometric
data in $B$, $V$, $R$ and $I$. While A is brighter by about 0.45\,mag in $I$
and $R$, the  
difference starts to decrease in $V$ and is compatible with zero in the
$B$ band.

Given the similarities of the spectra, the differences have to be
interpreted in a lensing context.
As discussed in the introduction, two possible explanations will be discussed
here. One is differential
reddening by different amounts of dust along both lines of sight in the
lensing galaxy, the other is microlensing which by different amplification
patterns for the two images in combination with a colour gradient of 
the source can also result in chromatic effects.

To separate these effects, we use the assumption that the regions
causing the 
emission lines are much larger than the continuum region so that the
microlensing signal is smeared out in the emission lines. We can thus
use the emission lines to study the pure extinction signal. In the next step
we correct the continuum (which is affected by extinction \emph{and}
microlensing) for extinction to study the microlensing effect.
To accomplish these tasks, we have to fit the continuum and determine the flux
ratios of the continuum-subtracted emission lines.

\begin{figure*}
\includegraphics[width=0.95\textwidth]{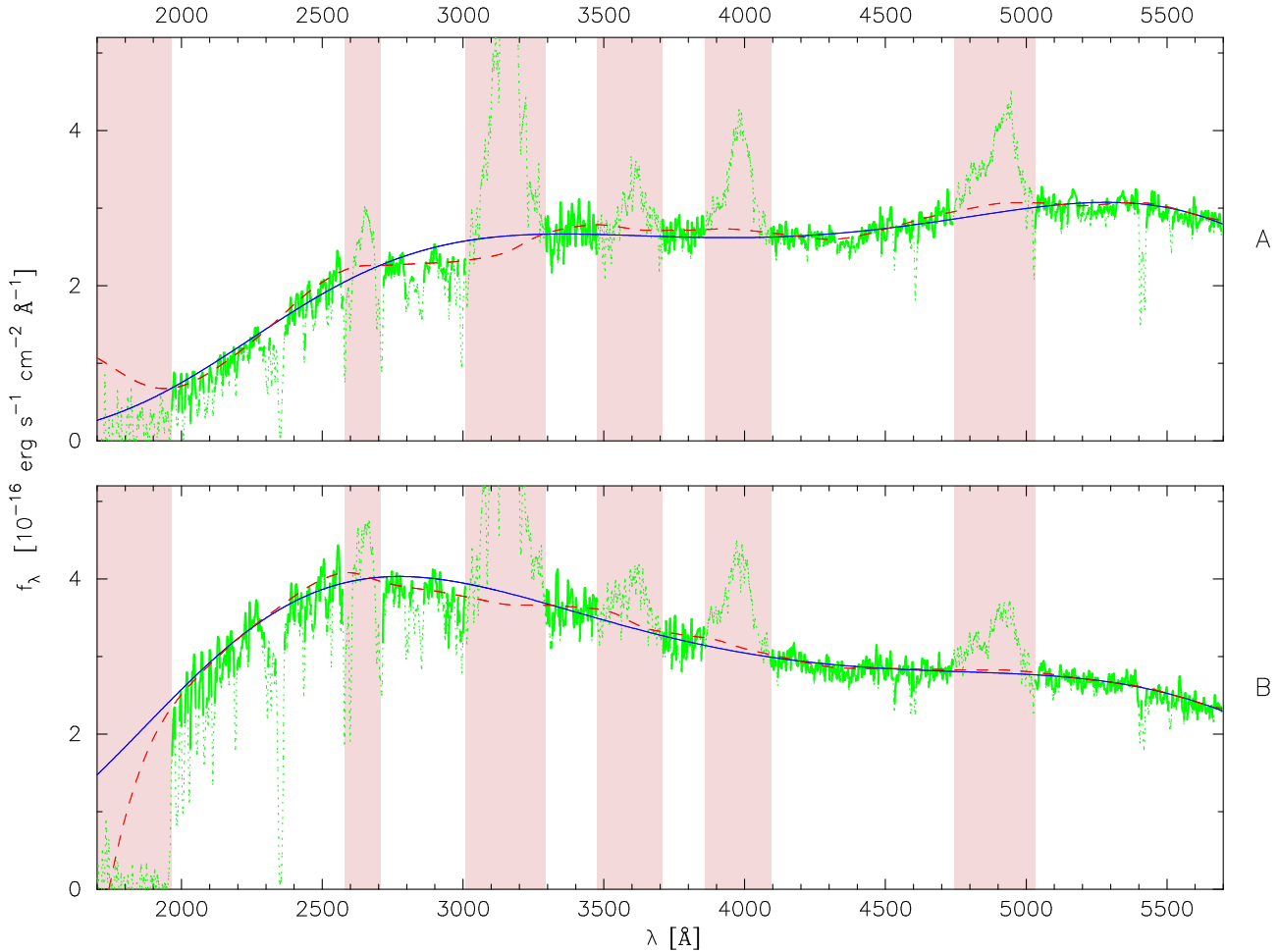}
\caption{Observed spectrum of A (top) and B (bottom) together with the two
  different continuum fits. The global polynomial fit is shown solid, the 
  locally smoothed estimate is dashed. The shaded regions mark the emission
  lines and the region affected by the Lyman limit on the far
  left. Parts of the 
  spectra not included in the fit (in the emission line regions or rejected
  by the global polynomial fitting algorithm because of low flux) are shown
  dotted, the rest of the spectra is solid.
}
\label{fig:cont}
\end{figure*}

\begin{figure*}
\includegraphics[width=12cm]{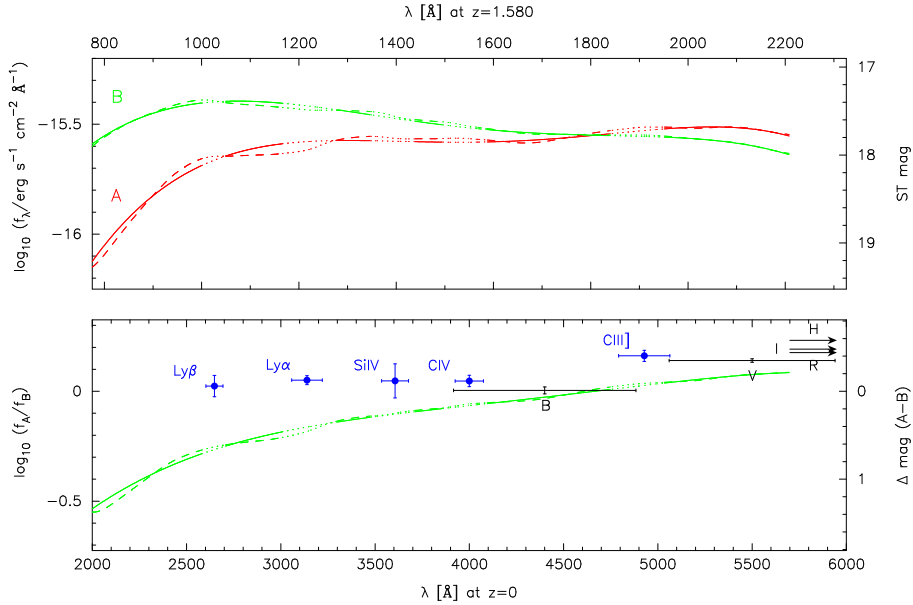}\hspace{1em}
\caption{Top: Continuum flux of A and B.
Bottom: Flux ratio $\mathrm{A}/\mathrm{B}$ (curve). The error crosses
  above the lines are for the continuum subtracted emission lines, where the
  horizontal bar shows the FWHM of the lines.
  Photometric measurements (combined continuum and emission lines) are shown at
  the right (see text).
  The photometric error bars are formal statistical uncertainties.
  $R$, $I$ and $H$ are outside of the wavelength range and are shown as arrows.
  The same line type coding as in Fig.~\ref{fig:cont} is used for the
  continuum fits in both panels. The regions
  of emission lines were not used for the fits and are shown dotted.
}
\label{fig:ratio}
\end{figure*}

\subsection{Global continuum fit}

We first flagged the regions of the five strong emission lines of both
spectra and kept only the continuum which is still covered with many
absorption lines.
We tried two approaches to fit the global continuum. One is a global
fit of a polynomial of degree 4 to the logarithmic flux. The other is
a smoothing 
algorithm which works by fitting a quadratic function to the spectrum for each
wavelength, using a Gaussian weighting with FWHM of 300\,\AA\ to effectively
include only nearby data points. The central value of this local fit is taken
as the result for the central wavelength and the procedure is repeated for
each wavelength. This algorithm
is more appropriate than a simple Gaussian convolution filter because it can
follow local slopes and curvatures more accurately. In particular it avoids
any edge effects.
In both algorithms we allowed for a certain fraction of pixels to deviate in a
negative direction from the model to avoid absorption lines affecting the
fit. These pixels were then discarded in the $\chi^2$ summation. We started
with this fraction set to zero and then increased it slowly until a level of
30\,\% was reached. This value was chosen to obtain a final normalized
$\chi^2$ close to unity. It is equivalent to a final effective
(one-sided) clipping limit of about 1.5--$1.7\,\sigma$.

The global continuum fits are shown in Fig.~\ref{fig:cont} together with the
observed spectrum.
In the comparison, the locally smoothed continuum follows the spectrum more
closely while the global polynomial fit is slightly smoother on large
scales. Both curves look like good continuum fits to the eye, and
the small differences are not significant for the further analysis which was
performed with both fits.
We have to keep in mind, however, that the continuum short-wards of the
\ion{Ly}{$\beta$} emission line might still be significantly contaminated by
the 
Lyman forest. This should not be a problem in the continuum subtraction of the
emission line, because the broad emission lines are affected by the Lyman
forest as well. The shape of the continuum near the short wavelength end itself
should nevertheless be interpreted with caution.
Figure~\ref{fig:ratio} shows the continuum of A and B (top) as well as
the ratio of the two (bottom) as a function of $\lambda$.

\subsection{Emission line fluxes}

The main problem in calculating the emission line fluxes is the continuum
subtraction. We used three different estimates for the continuum to get a
feeling for the accuracy. The first two are the global polynomial fit and the
smoothed continuum as discussed before. The third method uses local linear
trends with fitting regions on both sides and directly next to the emission
lines. 
After subtracting the continuum, different methods were used to measure the
emission line flux. The simplest one uses weighted averaging over two
different apertures, one narrow (27--59\,\AA) and one relatively wide
(74--261\,\AA). The widths of these windows were adapted `by eye' to the
individual emission lines. The narrow windows just enclose the core of the
lines while the wide ones are chosen in a way to include the complete emission
lines. We also performed
Gaussian fits and a local quadratical smoothing (as explained before, Gaussian
FWHM=100\,\AA) of the flux, also with wide and narrow fitting regions. The
values at the nominal line centre of these 
fits were then used as the maximal line flux.

A byproduct of the Gaussian fits is the QSO redshift. For the five
emission lines, we obtain a mean value of $z=1.582$ with a rms scatter of
$0.008$ which is dominated by the different line shapes. The statistical
error is much smaller, of the order 0.001. This compares moderately well with
the result 
from \citet{gregg00} for the combined spectrum of $z=1.565$ for \ion{C}{iv}
and \ion{C}{iii]}.
There is no noticeable difference in the results for A and B.

Figure~\ref{fig:ratio} (bottom) shows the resulting emission line flux
ratios compared with the continuum ratio. The plotted 
values are the mean results from our different estimates. As error bars we
used the formal statistical error added in quadrature to the internal scatter
of all methods to include possible systematics.
There seems to be almost no variation for the lines \ion{C}{iv}, \ion{Si}{iv},
\ion{Ly}{$\alpha$} and \ion{Ly}{$\beta$}. All of these are compatible with
their 
mean of $\log (f_\mathrm{A}/f_\mathrm{B})=0.04$.
Only for \ion{C}{iii]} we notice a deviation from the other values.
The photometric $BVRI$ measurements in that plot are from \citet{gregg00}, $H$
is 
from a deep 2400\,sec VLT/ISAAC exposure (details will be published
elsewhere).

\begin{figure*}
\includegraphics[width=12cm]{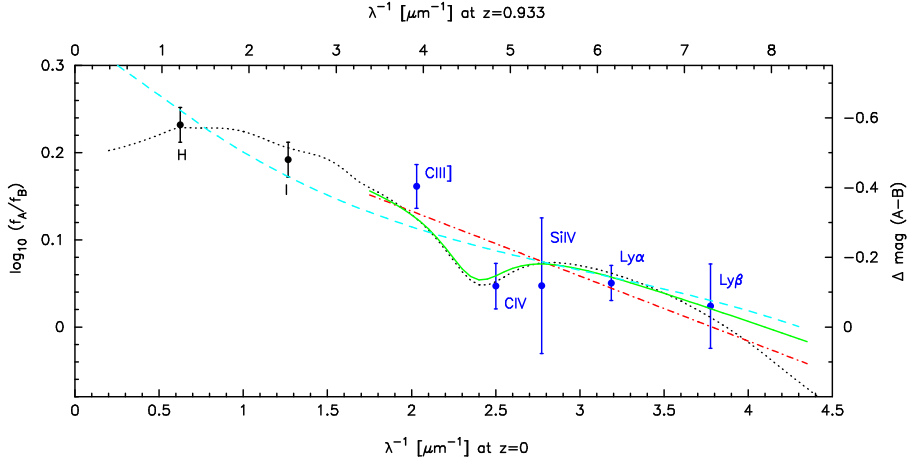}\hspace{1em}
\caption{Different extinction laws fitted to the data from emission lines and
  red/infrared broad band photometry. The photometric error bars were extended
  to $\pm0.05\,\mathrm{mag}$ (see text).
Models with 2175\,\AA\ bump: solid (FM bump) and dotted (CCM); models without:
dashed (\cab) and dot-dashed (FM nobump).
}
\label{fig:extinct}
\end{figure*}

Figure~\ref{fig:emlines} shows the emission line profiles after subtraction of
the local linear continuum fit and rescaling with the estimated emission
line flux ratio. Apart from subtle differences in absorption features, the
emission line profiles of A and B are remarkably similar.
The remaining
differences 
seem to be largest in $\ion{Si}{iv}$ but this is a result
of the weakness of this line which also leads to the relatively large
photometric uncertainty shown in Fig.~\ref{fig:ratio}.

The differences of the absorption left of
\ion{Ly}{$\beta$} are a result of the very different relative strength of the
absorption systems in A and B.
Altogether the profiles are highly consistent, leaving very little doubt
about the lensed nature of the two images.

The emission line ratios determined from the wide (complete line) and narrow
(line core) windows are also 
compatible with each other. This means that the microlensing signal is the
same for the broad and narrow line regions. Since the two are expected to be
very different in size, the effect would be different if there is any
effect at all. This means that microlensing seems to be absent
already for the broad lines. Using emission lines therefore seems a
reasonable way to measure extinction curves which are unaffected by
microlensing, at least in \he.

\section{Differential extinction}

To analyse the differential extinction, we fitted different empirical
extinction laws to the emission line data and the photometric data in $I$ and
$H$,
assuming that microlensing is affecting these bands only weakly as
indicated from the similarity of the extrapolated continuum and the
photometric data (Fig.~\ref{fig:ratio} bottom).
To take into account the residual microlensing effects and flux ratio
changes by variability within the time-delay, we used
an error of $\sigma=0.05$\,mag for the photometric data points instead of the
formal error or 0.02\,mag. The results are not very sensitive to this
  exact value.

The first extinction law is the galactic curve from \citet{ccm} which has a
2175\,\AA\ feature whose strength is correlated to the far-UV behaviour (CCM).
Alternatively we used the parametrisation from \citet{fm} once with the Drude
term describing the 2175\,\AA\ bump (FM bump) and once without it (FM
nobump). The strength of the 
bump is a free parameter independent of the global shape of the extinction
law, but we fixed its position and width ($x_0=4.6\,\mumi$,
$\gamma=0.95\,\mumi$).
We neglected the far-UV term
$c_4\,F(x)$ \citep[eq. 2]{fm} because it becomes important only at
  shorter wavelengths and is thus
only very weakly constrained. The inclusion would actually increase the
normalized $\chi^2$ of the fits. The strength of the 2175\,\AA\ feature does
not depend on the far-UV term.
Although the FM models are valid only for $\lambda^{-1}\gtrsim 3.3\,\mumi$, we
included the $H$ and $I$ data in the fit using an extrapolation. The FM plots
should therefore only be seen as illustrations and not as 
physical models. Without $H$ and $I$ these curves would rise on the left to
match 
\ion{C}{iii]} more closely, without changing the 2175\,\AA\ bump very
  much.
As a second extragalactic model we used the
functional form from \citet{calzetti00} (\cab) without a 2175\,\AA\ feature.
All extinction law fits are shown in Fig.~\ref{fig:extinct}.

We see that all of these models fit the data relatively well; the normalized
$\chi^2$ values are all close to unity. To follow the slightly deviating
value from the \ion{C}{iv} line more closely, however, the 2175\,\AA\ bump has
to 
be included which also shows in improved residuals for the CCM and FM bump
models. Since the \ion{C}{iv} line lies in a region of the
spectrum were the 
continuum is very well defined, there is no reason to trust this data point
less than the others. Our result can therefore be seen as a marginal
detection of the 2175\,\AA\ feature in the spectrum of \he. This
interpretation is supported by the fact that the free fit of the absorption
feature in the FM bump model favours a strength almost exactly equal to the
galactic CCM model.

For the physical interpretation we only consider the galactic CCM law
\begin{equation}
\frac{A(\lambda)}{A(V)} = a(\lambda^{-1}) +
\frac{b(\lambda^{-1})}{R_V} 
\end{equation}
with $R_V=A(V)/E(B-V)$. The functions $a$
and $b$ are defined piecewise over a region $0.3\,\mumi \le \lambda^{-1} \le
10\,\mumi$ \citep{ccm}.
For the differential extinction law for $\mathrm{A}-\mathrm{B}$ we obtain
\begin{equation}
A\supp{A}(\lambda)-A\supp{B}(\lambda) = p_a \, a(\lambda^{-1}) + p_b \,
b(\lambda^{-1}) 
\end{equation}
with the two independent parameters
\begin{eqnarray}
p_a &=& A\supp{A} (V) - A\supp{B} (V) \\
p_b &=& \frac{A\supp{A} (V)}{R\supp{A}_V} -
 \frac{A\supp{B} (V)}{R\supp{B}_V} \rtext{.}
\end{eqnarray}
The first parameter $p_a$ is the differential visual extinction, the second
one is the 
differential colour excess $p_b=E\supp{A} (B-V) -
E\supp{B} (B-V)$.
The difference of measured magnitudes is equal to the difference in intrinsic
brightness (caused by different amplification and variability in combination
with the time-delay) $p_0$ plus the difference in extinction.
We therefore have three free parameters.

\begin{table}
\centering
\caption{Parameters of CCM model fits in magnitudes. $p_0$ is the `true' brightness
  difference, corrected for extinction but including the different lens
  amplifications. $p_a$ is the differential extinction in $V$, 
  $p_b$ is the differential colour excess. The effective $R_V$ is
  calculated as $p_a/p_b$. It is the value of $R_V$ if the shape of the
  extinction laws in A and B would be equal or if extinction is affecting only
  one component. The first two lines show a free fit including $1\,\sigma$
  error bars, $R_V\supp{eff}$ was fixed for the 
  other lines.}
\label{tab:par}
\begin{tabular}{lllll}
$p_0$ & $p_a$ & $p_b$ & $R_V\supp{eff}$ & $\phantom{0}\chi^2/\nu$ \\ \hline
$-0.493$ & $-0.074$ & $+0.037$ & $-2.00$ & $\phantom{0}3.2/4$ \\
\multicolumn{1}{r}{$\pm0.11\phantom{0}$} & \multicolumn{1}{r}{$\pm0.11\phantom{0}$} &
\multicolumn{1}{r}{$\pm0.015$} &  \multicolumn{2}{r}{(free fit)}  \\ [1.5ex]
$-0.672$ & $+0.131$ & $+0.059$ & $+2.20$ & $\phantom{0}6.1/5$ \\
$-0.733$ & $+0.206$ & $+0.067$ & $+3.10$ & $\phantom{0}8.8/5$ \\
$-0.870$ & $+0.447$ & $+0.077$ & $+5.80$ & $34.3/5$
\end{tabular}
\end{table}

\begin{figure}
\centering
\includegraphics[width=\hsize]{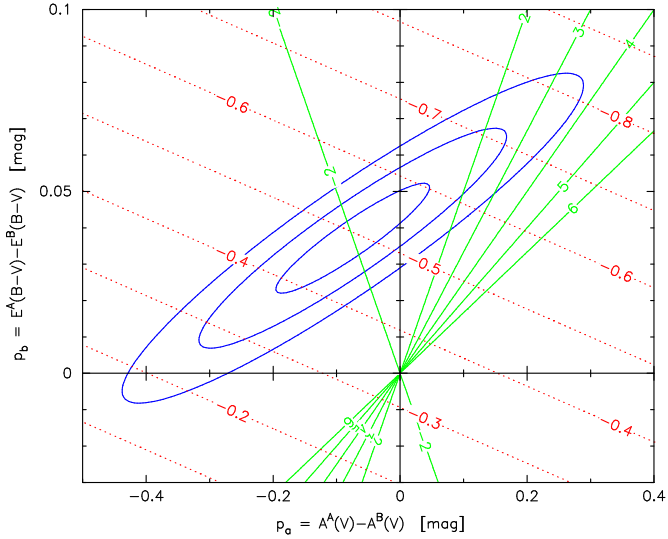}
\caption{Residuals for fixed parameters $p_a$ and $p_b$ of the CCM model. The
  ellipses show differences of $\chi^2$ from its minimal value of $1$, $4$ and
  $9$ corresponding to confidence limits of $1$, $2$ and $3\,\sigma$ for one
  parameter. The solid lines intersecting at $(0,0)$ show the effective
  $R_V=p_a/p_b$. We see that the data are compatible with one value of
  $R_V$ only if this value is very low. The best fit has $p_a/p_b=-2.00$.
The dotted parallel lines show the fitted $p_0$ which gives the true
  amplification ratio of the images.
}
\label{fig:extchi2}
\end{figure}

The parameters of the best fit are shown in Table~\ref{tab:par}. We see that
the visual extinction $A(V)$ has to be higher in B by 0.07\,mag but
the reddening $E(B-V)$ is higher in A by 0.04\,mag. Recall
that $A(V)$ and $E(B-V)$ are not measured directly in $V$ and
$B$ but are merely used as parameters to describe and fit the UV extinction
law. 
Figure~\ref{fig:extchi2} shows residuals of fits with different fixed values of
$p_a$ and $p_b$. The data are only marginally compatible with only one value
of $R_V$ (same value in both images or extinction in only one of the
images) in the realistic region of $2.2<R_V<5.8$. The best effective 
$R_V=p_a/p_b$ is $-2.00$ which can only be explained with a combination of
extinction both in A and B with different $R_V$.
To explain the measurements with the smallest possible visual extinction,
$R\supp A_V$ must be at its lower and $R\supp B_V$ at its upper
limit. For $R\supp A_V=2.2$ and $R\supp B_V=5.8$ we get $A\supp
A(V)=0.177\,\mathrm{mag}$ and $A\supp B(V)=0.251\,\mathrm{mag}$. For more
similar extinction laws, the 
total extinction has to be higher, e.g.\
$A\supp A(V)=0.668\,\mathrm{mag}$ and $A\supp B(V)=0.742\,\mathrm{mag}$ for
$R\supp A_V=3$ and $R\supp B_V=4$. Such values are not atypical for
galactic extinction. Even with more similar $R_V$ the total extinction
does not become unrealistically high.

The highly significant differences between continuum and emission line ratios
(Fig.~\ref{fig:ratio}) show the importance of microlensing when studying
differential extinction curves. Using photometric data points without
correcting for microlensing like in \citet{falco99} would in the case of \he\
lead to highly incorrect results.

An important result for lens models is the true amplification ratio which
corresponds to $p_0=\mathrm{A}-\mathrm{B}=-0.49\pm0.11\,\mathrm{mag}$ ($1\,\sigma$ error
bar). The amplification ratio is thus $|\mu\sub A/\mu\sub B|=1.57\pm0.17$. Note
how different this value is from the continuum or emission line
ratio in the UV (Fig.~\ref{fig:ratio}). The combination of extinction and
microlensing changes the continuum flux ratio at the short wavelength end by
almost 2 magnitudes. One should therefore be very careful when using optical
flux ratios as constraints for the amplification ratio in lens modelling if no
correction for microlensing and differential extinction can be applied.
The result for $|\mu\sub A/\mu\sub B|$ depends on the emission lines
only. Variations of the \emph{continuum} during the time-delay (see estimate
below) 
would not change this ratio. Variations of the emission lines are still a
possible source of error here but are expected to be much smaller because of
the much larger size of the emitting regions.

\section{The lensing galaxy}

The available observational data are quite insufficient to constrain
realistic 
lens models to a satisfying degree. Since not even an estimate of the galaxy
position is available, the best we can do at the moment is to use the simplest
model, a singular isothermal sphere without external perturbations, and
estimate the mass with this approach. In contrast to the time-delay,
the mass depends only weakly on the exact galaxy position or asymmetries of
the mass distribution.

With the source redshift $z\sub s=1.58$ and a proposed lens redshift $z\sub
d=0.933$,
this leads to an expected radial velocity dispersion of
$\sigma_v=192\,\mathrm{km\,s^{-1}}$ and an
enclosed mass within the Einstein radius of $M\sub
E=6.8\cdot10^{10}\,M_{\sun}$.
Using the Tully-Fisher relation for late-type galaxies at intermediate redshift
\citep{ziegler02}, we obtain an absolute luminosity of $M_B=-22\,\mathrm{mag}$
using an equivalent circular velocity of $v_c=330\,\mathrm{km\,s^{-1}}$. This
corresponds to an apparent $I$ magnitude of $21.5\,\mathrm{mag}$, about
4.5\,mag weaker than each QSO component. Direct images taken with STIS in the
same campaign do not show any obvious sign of the core of the lensing galaxy
after applying a standard subtraction of the two QSO images. We are currently
working on new methods to optimize this PSF subtraction process.

The maximal time-delay would be reached if the lens is located very close to
one of the images, leading to a limiting value of 170\,days. For this scenario
a very high ellipticity of the mass distribution would be required. More
realistic values are $\sim\,$1--2 months. A spherically symmetric lens with a
true 
amplification ratio of 1.57 (the result of the previous section and assuming
negligible variability) would lead to a time-delay of $\Delta t=38\,$days with
A leading. Even a moderate ellipticity or external shear could change this
number (even its sign) considerably.

\section{Microlensing}

\begin{figure}
\centering
\includegraphics[width=\hsize]{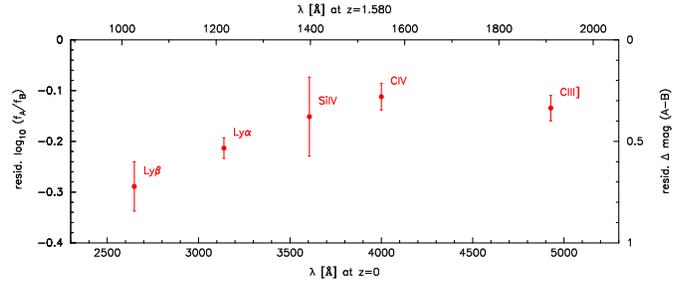}
\caption{Differential microlensing amplification ($A/B$) as a function of
  wavelength (error bars). The continuum (affected by microlensing and
  extinction) was 
  corrected for extinction using the emission line fluxes which are expected
  to be unaffected by microlensing.
}
\label{fig:micro}
\end{figure}

With the measurement of the emission line flux ratios, we were able to obtain
estimates of the differential extinction. Since the continuum is affected by
extinction \emph{and} microlensing, we have to remove the extinction effect to
study the microlensing signal alone.
Figure~\ref{fig:micro} shows the continuum flux ratio corrected for
differential 
extinction as determined from the emission lines.
Given the high uncertainties in estimating the extinction law, we do not
show the estimated global microlensing function but only the
values for the wavelengths of the emission lines which have been measured
directly and which are thus not model dependent.

These data points should be flat zero in the absence of microlensing. We
see that in reality either B is 
amplified and/or A is de-amplified with a strength increasing for
smaller wavelengths.

An additional complication is the possibility of variability. Because of
the time-delay, the two components are observed in two different epochs in
the source plane. If the source brightness or spectrum changes significantly
over this time-delay, the measured ratio would not reflect the true ratio
anymore.
Assuming that this effect is negligible in our observations of \he, we can
use the magnitude of the differential microlensing to estimate an upper
limit for the source size in terms of the Einstein radius using the recipe
of \citet{rs91}.
If the observed differential microlensing signal of $\delta
m=0.5\,\mathrm{mag}$ is typical for the microlensing variance, a
conservatively estimated upper limit of 4 Einstein radii can be derived.
The Einstein radius in our case is $\theta\sub e =1.2\,\muup\mathrm{arcsec}
\cdot\sqrt{M/M_{\sun}}$ for microlenses of mass $M$,
corresponding to $0.01\,\mathrm{pc}$ ($2200\,$AU) in the source plane for
$M=M_{\sun}$. 

We also tried to fit the parameters of more specific physical scenarios like
isolated 
caustic crossings of a standard accretion disk model to the data, only to
learn that (although this should be a powerful method in principle) the
available data are not sufficient to obtain any conclusive results.

\section{Summary}

We presented the first separate spectra of both components of the doubly
lensed QSO 
\he. The similarity of the optical and near UV spectra
strongly suggest the lens nature of this object.

The known optical colour differences of the images, which (together with the
metal absorption 
lines) were the main motivation to start this project, extend to the near UV
and 
become even stronger in this range. While A is brighter than B by about
0.5\,mag 
in $R$ and $I$ band, this relation is more than inverted in the continuum near
$2000\,\AA$ where B is brighter by about 1.3~mag.

We used the continuum subtracted emission line flux ratios to
study the differential extinction in both images. The line emitting regions of
the QSO are confirmed to be so large that any residual microlensing effects on
their fluxes is insignificant so that the extinction can be studied
unambiguously.
We fitted several empirical extinction laws from the literature and found that
curves with a significant 2175\,\AA\ bump (CCM: \citealp{ccm}; FM bump:
\citealp{fm}) reproduce the data better than typical extragalactic curves
without this feature 
(FM nobump: \citealp{fm}; \cab: \citealp{calzetti00}).
Although the differences are not highly significant, this provides at least
evidence 
that the 2175\,\AA\ feature can also be present in high redshift galaxies
\citep[compare][]{motta02}.

From the difference between the emission line and continuum ratios we learned
that microlensing contributes significantly to the variations of flux ratio
with wavelength. On the short wavelength end, the microlensing signal is
actually much stronger than the differential reddening. While the emission
lines have almost the same flux in A and B there, the continuum is
brighter in B by up to 1.3\,mag (see
Fig.~\ref{fig:ratio} bottom). For our study it was therefore absolutely
essential to use the emission line fluxes for the extinction analysis.
Analysing 
the continuum or broad band photometry like in \citet{falco99} without
correcting for microlensing would have led
to highly incorrect results in the case of \he.

The analysis of galactic extinction law fits showed that the data are only
marginally consistent with the same extinction law in A and B with only
different strength. The best fit has a higher total visual extinction $A(
V)$ in B but a higher reddening $E(
 B-V)$ in A.

Using the image separation, we were able to estimate the mass within the
Einstein radius to be $M\sub E=6.8\cdot 10^{10}M_{\sun}$. 
The expected
radial velocity dispersion of an isothermal model is
$\sigma_v=192\,\mathrm{km\,s^{-1}}$, equivalent to a circular velocity of
$v_c=330\,\mathrm{km\,s^{-1}}$.
The time-delay is expected to be of the order 1--2 months.

In the last part we used the differential extinction measurement from the
emission lines to correct the continuum flux ratio at five wavelengths
for this effect.
The remaining ratio is a direct measurement of the microlensing
signal alone.
This is the first time that such a study is carried out to
separate the two effects and discuss them individually.

The microlensing data are not sufficient to determine the parameters of
physically motivated source models accurately but could be used to estimate an
upper limit for the effective source size using the assumption
that variability over a scale of the time-delay does
not change the observed flux ratio significantly.
To test this assumption and to determine the time-delay, photometric
monitoring of this system should be performed in the future.
The information gained by our work 
(spectral microlensing measurements
at one epoch) is complementary to the more common light-curve analysis
technique (photometric measurements at many epochs, see e.g.\
\citealp{shalyapin02} for recent results).
A combination of both methods by using spectrophotometric monitoring data has
the potential to lead to more valuable constraints than both
methods alone by breaking degeneracies inherent in the individual
approaches. It will then be possible to determine brightness profiles in
all of the measured wavelengths which can be combined to a
relatively model-independent measurement of the source's temperature profile,
allowing to test different physical QSO models.

\begin{acknowledgements}
The authors like to thank P.~Schechter for helpful discussions.
This work was supported by the Verbundforschung under grant 50~OR~0208.
SL acknowledges support from the Chilean  
{\sl Centro de Astrof\'\i sica} FONDAP No. 15010003, and from  FONDECYT 
grant N$^{\rm o} 3\,000\,001$.

\end{acknowledgements}


\end{document}